\documentclass[lettersize,journal]{IEEEtran}

\usepackage[hyphens]{url} 
\usepackage{hyperref}     
\usepackage{amsmath,amssymb,amsfonts}
\usepackage{mathtools}
\usepackage{algorithmic}  

\usepackage{cite}
\usepackage{float}
\usepackage{graphicx}
\usepackage{textcomp}

\usepackage{algorithm}
\usepackage{array}
\usepackage[caption=false,font=normalsize,labelfont=sf,textfont=sf]{subfig}
\usepackage{textcomp}
\usepackage{authblk}
\usepackage{stfloats}
\usepackage{adjustbox, diagbox, caption}
\usepackage{tabularx, multirow, multicol, makecell}

\begin{document}

\title{Speech Separation for Hearing-Impaired Children in the Classroom}
\author[1,*]{Feyisayo Olalere}
\author[1,2]{Kiki van der Heijden}
\author[3]{H. Christiaan Stronks}
\author[3]{Jeroen Briaire}
\author[3,4,5]{Johan H. M. Frijns}
\author[1]{Yagmur Güçlütürk}
\affil[1]{Radboud University, Donders Institute for Brain, Cognition, and Behaviour, The Netherlands}
\affil[2]{Mortimer B. Zuckerman Mind, Brain, Behavior Institute, Columbia University, USA.}
\affil[3]{Department of Otorhinolaryngology, Leiden University Medical Centre, The Netherlands}
\affil[4]{Leiden Institute for Brain and Cognition, Leiden, The Netherlands}
\affil[5]{Department of Bioelectronics, Delft University of Technology, Delft, The Netherlands}
\affil[*]{corresponding: feyisayo.olalere@donders.ru.nl}
\maketitle

\begin{abstract}
Classroom environments are particularly challenging for children with hearing impairments, where background noise, simultaneous talkers, and reverberation all degrade speech perception. These difficulties are even greater for hearing-impaired children than for adults, yet most deep learning–based speech separation algorithms for assistive devices are developed for adult voices in simplified, low-reverberation conditions. This neglects both the higher spectral similarity of children’s voices, which reduces separation cues, and the acoustic complexity of real classrooms.
We address this gap by applying MIMO-TasNet, a compact, low-latency, multi-channel speech separation architecture well-suited for real-time processing in bilateral hearing aids or cochlear implants. Using this model, we simulated naturalistic classroom conditions with moving talkers, both child–child and child–adult pairs, across varying noise and distance settings. The training strategies were designed to test how well the model could adapt to children’s speech by leveraging spatial cues. We trained on adult speech only, a simulated classroom dataset, and finetuned the adult-trained model with a smaller subset of the same classroom data to assess data-efficient transfer learning.
Our results show that with the right spatial cues, models trained solely on adult speech can still perform well in clean classroom environment, even on overlapping child talkers. Training on classroom-specific data substantially improved separation quality, while fine-tuning with only half the classroom data achieved even greater performance, confirming the benefit of data-efficient adaptation. Furthermore, training with diffuse babble noise further improved robustness across clean and noisy conditions, and the model preserved spatial awareness of talkers while generalizing well to unseen talker–listener distances. These results show that combining spatially aware architectures with targeted, data-efficient adaptation strategies can significantly improve speech accessibility for children in noisy classroom environments, paving the way for practical, on-device assistive technologies.

\end{abstract}


\section{Introduction}
\label{sec:intro}

Classroom environments pose significant challenges for speech perception, especially for younger children \cite{klatte2013does}. These challenges are even more pronounced for children with hearing loss in classrooms designed for typical-hearing students \cite{neuman2010combined, gremp2018descriptive}. Younger children need more favorable signal-to-noise ratios than adults to understand speech in reverberant conditions \cite{neuman2010combined}, and modest background noise can disrupt their learning \cite{mcmillan2016learning}. Children with hearing loss, whether mild, unilateral, or using hearing devices, require better listening conditions to achieve comparable performance \cite{lewis2016effects, ching2018factors}. For those with auditory processing difficulties, noisy classrooms can further interfere with learning and participation \cite{jutras2019listening}. Overlapping talkers, reverberation, background noise, and talker-to-listener distance all contribute to the communication barriers faced by hearing-impaired children \cite{valente2012experimental,attia2025cpt}.

Hearing technologies like hearing aids and cochlear implants (CI) are essential assistive devices commonly used by hearing-impaired children \cite{henry2021noise}. While these devices offer substantial benefits in quiet, controlled settings, their performance often degrades in challenging real-world environments such as classrooms \cite{qin2003effects,healy2016difficulty,gazibegovic2024hearing}. Advanced hearing aids use technologies like beamforming to enhance speech from a speaker positioned directly in front or from the side of the listener \cite{mccreery2012evidence,gazibegovic2024hearing}. Although these systems have shown promising results in adults \cite{henry2021noise,de2024envelope}, their effectiveness is limited in school settings, where children often move unpredictably, speak from varying positions, and produce less structured, more spontaneous speech than adults \cite{wolfe2017evaluation,wolfe2022evaluation}.

Given these limitations, there is growing interest in supplementary approaches that enhance the speech signal for use in hearing aids and CIs \cite{johnstone2018using}. One such approach involves wireless remote microphone (WRM) systems \cite{gabova2024wireless}. In this setup, the talker wears or places a microphone nearby, which transmits audio directly to the listener’s hearing aid. WRMs have been shown to improve speech understanding, particularly during teacher-led instruction, by improving the signal-to-noise ratio \cite{chen2021effects,thompson2020remote}. However, this solution is less practical for children, who must also understand their peers during discussions and social interactions. Since peers typically do not wear WRMs, these systems fall short of addressing the full scope of children's communicative needs in classrooms \cite{gustafson2021individual,gabova2024parents}.

Recent advancements in artificial intelligence (AI), particularly in DNN-based speech separation, have opened up new possibilities for improving speech intelligibility in noisy and complex auditory environments\cite{zhao2024mossformer2,lee2024boosting,lutati2022sepit,subakan2021attention}. When used as a front-end for hearing assistive devices \cite{han2020real}, these models can isolate individual speech components from mixtures, allowing clearer access to the target speaker. In this approach, sound is captured by hearing aid microphones and processed by the DNNs in real-time to separate speech from noise or interfering speech, and the enhanced signal is returned to the listener. This method has the potential to eliminate the need for multiple WRMs, which is impractical in peer-to-peer classroom settings. Moreover, DNN-based models can adapt to dynamic listening scenarios, tracking speakers even as they move \cite{han2021binaural}. However, most state-of-the-art speech separation models have been developed and evaluated primarily on adult speech, with little to no testing on children's voices, despite known acoustic and linguistic differences. that affect model performance in related domains like speech recognition.

A consistent finding across speech recognition and enhancement studies is that models trained on adult voices generalize poorly to children’s speech \cite{southwell2024automatic,shivakumar2020transfer,attia2025cpt}. This performance gap is attributed to fundamental acoustic differences, such as higher fundamental frequencies, shorter vocal tracts, and greater intra-speaker variability in children’s speech. Importantly, prior studies that document this mismatch have almost exclusively relied on monaural recordings, where enhancement depends heavily on spectral distinctiveness. In such settings, the adult-to-child spectral mismatch produces a marked drop in performance. The potential for binaural cues such as interaural phase (IPD) and level differences (ILD) to mitigate this gap remains underexplored.

Binaural hearing has long been recognized as critical for parsing multi-talker environments, particularly in noisy or reverberant spaces \cite{bronkhorst2000cocktail,litovsky2006bilateral}. Our prior work also demonstrated that binaural models can leverage spatial cues to separate spectrally similar adult talkers \cite{olalere2025leveraging}. We hypothesize that since spatial cues can reduce reliance on spectral distinctiveness, they could thereby bridge part of the adult-to-child mismatch generally observed when using adult-trained speech enhancement models for children-related speech. We ask whether binaural information can also help an adult-trained speech separation model generalize to children’s speech. At the same time, we recognize that mismatches in noise type, such as adult versus child babble or having two child talkers overlap, where spectral similarity is greatest, are likely to remain challenging.

To address these challenges, this study focuses on speech separation in naturalistic classroom scenarios. We simulated these environments using binaural room impulse responses (BRIRs) from children's ears, capturing both listener- and room-specific acoustic properties. Our dataset consists of dynamic scenes with moving talkers and background babble. For the speech separation model, we used the MIMO-TasNet model \cite{han2021binaural}, a compact and low-latency architecture suitable for hearing aids, which can process binaural input and preserve spatial information over time. Capturing and preserving these trajectories is part of the aim of this study, as it enables the evaluation of motion tracking in naturalistic classroom conditions where talkers frequently change position. Using this model and our simulated datasets, we evaluated three training strategies: (i) models trained solely on adult speech mixtures (ii) models trained directly on our classroom dataset containing child–child and child–adult speech mixtures and (iii) finetuned models, initialized from adult-only training and adapted with 50\% of the classroom dataset. All three strategies were trained with and without additional babble noise.

This study makes four contributions. First, we show that binaural cues substantially reduce the adult-to-child speech enhancement mismatch in clean conditions, contrasting with prior speech tasks done with monaural audio (e.g., in speech recognition or enhancement) where an adult-to-child generalization gap is consistently observed \cite{shivakumar2020transfer, southwell2024automatic}. Second, we show that in the presence of children babble noise, training with child-like babble material is needed. Third, we establish fine-tuning as an efficient adaptation strategy, outperforming full classroom training while requiring only half the data. Finally, we evaluate spatial preservation through direction-of-arrival (DoA) error and test generalization to unseen talker distances, highlighting the practical relevance of spatially aware models for hearing-assistive technologies. Together, these experiments allowed us to characterize how child-specific acoustic and spatial properties affect speech separation performance and to establish a foundation for deploying such models in ecologically valid environments like classrooms.

The remainder of this paper is structured as follows: in Section II, we describe the method, which contains the problem definition, and the simulation of overlapping speech in naturalistic classroom settings, including spatial modeling of moving talkers. Section III outlines the experimental framework, including model training procedures and evaluation metrics. In Section IV, we present the results of our evaluations, highlighting the performance gap between adult-trained models and classroom overlapping speech scenarios. We also show the improvements gained through fine-tuning and robust data training. Section V discusses the broader implications of our findings for speech enhancement in classroom environments, particularly for children using hearing aids. Finally, Section VI concludes the paper and outlines directions for future work.

\section{Method}

\subsection{Problem Definition}

The goal of this study is to perform speech separation in naturalistic classroom scenarios involving two simultaneously, moving talkers. These talkers are either two children or a child and an adult, simulating typical classroom scenarios involving overlapping speech between peers or between a student and a teacher.

Formally, given a binaural mixture signal 
$y_{s_1 + s_2} \in \mathbb{R}^{C \times T}$, 
where $C$ is the number of channels (here $C = 2$, representing sound at the left and right ears) and $T$ is the number of time samples, the task is to estimate the individual speech signals $\hat{s}_1$ and $\hat{s}_2$ corresponding to each active talker. We use binaural modeling to replicate the signals that would be captured by the microphones of a child with bilateral hearing aids, where spatial cues from each ear are crucial for separating talkers. In our scenario, the talkers move during the utterance, producing dynamically changing spatial cues that the model must track to maintain separation performance. This motion tracking is a central aim of our setup, as it reflects naturalistic classroom listening conditions in which talkers frequently change position. 

The input waveform $\mathbf{y}$ may also include additive child babble noise in the background, making the separation task more challenging and reflective of classroom acoustics. The separation model is trained to output time-domain waveforms for each speaker with no interference from the other speaker or background noise. This setup enables the model to not only separate overlapping speech, but also to preserve spatial distinctions associated with each moving source.

\subsection{Simulation of Overlapping Speech for Classroom Conditions}
\label{sec:simulation_overall}
\begin{figure*}[!t]
  \centering
    \includegraphics[width=\textwidth]{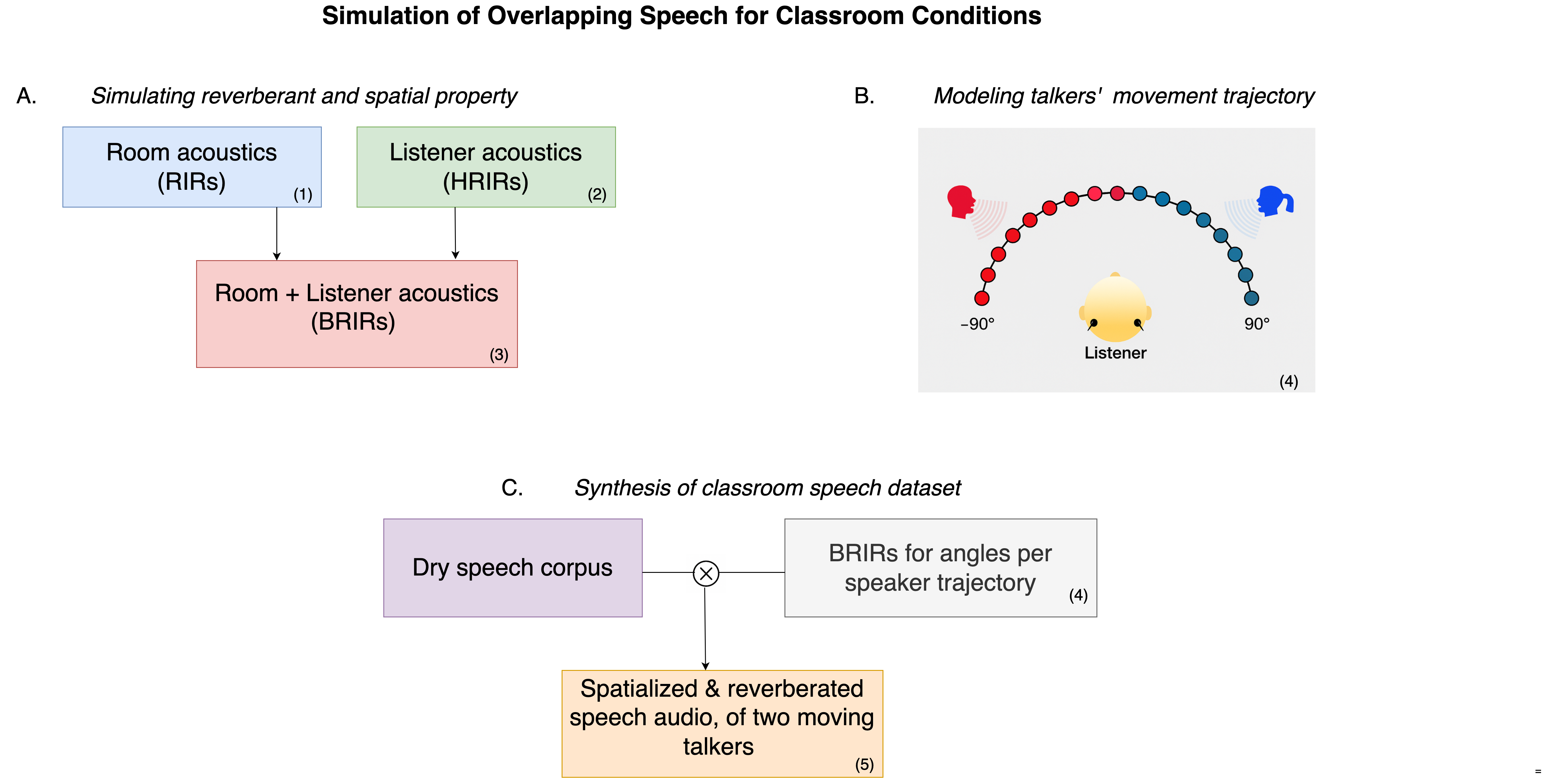}
    \caption{Simulation pipeline illustrating the generation of reverberant and spatialized speech mixtures for classroom conditions. The process includes simulating room and listener acoustic properties (A), modeling talkers’ movement trajectories (B), and synthesizing classroom speech mixtures (C). The numbers (1) - (5) correspond to the steps itemized in section \ref{sec:simulation_overall}}
    \label{fig:pipeline}
\end{figure*}

To capture the reverberant and spatial characteristics typical of classroom environments, we developed a spatialization pipeline for generating training and evaluation data (see Fig.\ref{fig:pipeline}). This pipeline consists of five main components, which are explained below in detail:

\begin{enumerate}
    \item Simulation of room impulse responses (RIRs)
    \item Application of head-related impulse responses (HRIRs)
    \item Generation of binaural room impulse responses (BRIRs)
    \item Modeling of talkers' movement trajectories
    \item Synthesis of the classroom speech data
\end{enumerate}


\subsubsection{Room Impulse Responses}
\label{rirs}
\noindent To simulate naturalistic reverberant classroom acoustics, we generated RIRs that capture direct sound, early reflections, and reverberation or echo. These RIRs were used to spatialize source signals in simulated classroom environments with varying geometry, reverberation, and source–listener distances. We used the Pyroomacoustics Python package \cite{scheibler2018pyroomacoustics}, which implements the image source method to model sound propagation in rectangular (shoebox) rooms.

A total of 30 classrooms were simulated, with dimensions randomly sampled from a range of $8.5 \times 8.5 \times 3$ m to $10 \times 10 \times 3.5$ m (length × width × height), reflecting typical U.S. classroom sizes \cite{crandell2000classroom,american2010acoustical}. Reverberation times ($T_{60}$) were drawn uniformly from 0.2 s to 0.7 s in 0.1 s increments to model variability in acoustic conditions found in real educational settings \cite{larson_davis_rt60,acoustic_frontiers_rt60}.

Listener positions were randomized on a 1-meter spaced grid in each room, ensuring a minimum distance of 1 meter from any wall to avoid boundary effects. For each simulation, one listener position was randomly selected from this grid to introduce spatial variability across the dataset. At each of these positions, we simulated RIRs as captured by a virtual multi-microphone array. This array consisted of six DPA-4060 omnidirectional microphones arranged in orthogonal pairs along a 10 cm diameter circle, with an Earthworks M30 or M50 microphone placed at the center. 

Furthermore, we simulated varying talker–listener distances by defining circular trajectories around each listener at radii of 1.0 m, 1.5 m, and 2.0 m. Along each circle, 72 talker positions were defined in 5$^\circ$ steps, resulting in 2,160 RIRs per distance setting (72 positions × 30 rooms). Importantly, the RIRs corresponding to 1.5 m and 2.0 m radii were used only at test time, allowing us to evaluate the model’s ability to generalize to talker–listener distances not seen during training.





\subsubsection{Head-Related Impulse Responses}

\noindent HRIRs encode the direction-dependent filtering effects of the outer ear, head, and torso on incoming sounds, providing essential cues for spatial hearing \cite{gardner1994hrft}. Unlike generic or KEMAR-based HRIRs \cite{burkhard1975anthropometric}, listener-specific HRIRs enable more accurate binaural rendering. Because our focus is on supporting hearing-impaired children, all mixtures in the main experiments were rendered from the child-listener's perspective. For this, we used the CHASAR HRIR dataset \cite{braren2021towards}, which includes measurements from children aged 5–10 years with microphones embedded in silicone hearing-aid domes, sampled on a 5-degree azimuth and elevation grid.

The CHASAR dataset was recorded with sources placed 1 meter from the receiver. Training on this fixed distance allowed us to control for variability in spatial cues and to explicitly assess generalization to unseen distances (1.5 m and 2.0 m), which were simulated by applying inverse-square law amplitude corrections. While this approximation does not model distance-dependent spectral effects (e.g., air absorption or early reflections), it provides a reasonable proxy for energy-level differences in far-field conditions \cite{blauert1997spatial}.


\subsubsection{Binaural Room Impulse Responses}
\label{BRIR}

\noindent BRIRs combine room-specific acoustic characteristics captured by RIRs with listener-specific spatial filtering effects modeled by HRIRs \cite{gari2019flexible}. Convolving a single-channel sound signal with a BRIR yields a spatialized, reverberant binaural signal that reflects both the room environment and the listener's anatomical filtering.

To generate these two-channel BRIRs, we adopted the methodology detailed in the study by Amengual et al. \cite{amengual2021optimizations}. This approach leverages the Spatial Decomposition Method (SDM) \cite{tervo2013spatial} to efficiently convert the multi-channel RIRs (generated from our virtual 7-microphone array, as described in Section \ref{rirs}) into directional information, which is then convolved with the listener-specific HRIRs. This process projects the spatial sound field, captured by the microphone array, onto the two ear channels of a binaural listener. This multi-stage process was applied to all simulated RIRs at each listener–talker distance, resulting in a total of 2,160 BRIRs per distance condition (72 directions × 30 rooms), each being a two-channel (left and right ear) impulse response.

\subsubsection{Modeling Movement Trajectories}
\label{trajectories}

To simulate spatially dynamic speech sources, we adapted the moving speaker simulation approach detailed in the study by Han et al. \cite{han2021binaural}. In this framework, each talker moves along a trajectory constrained to the frontal azimuthal plane, spanning from -90$^{\circ}$ to 90$^{\circ}$ with a 5$^{\circ}$ angular resolution, consistent with the spatial resolution of our BRIRs.

Each talker's trajectory is dynamically determined over the 2.4-second utterance length. They begin at a randomly selected azimuth angle, and their movement direction (clockwise or counterclockwise) is randomly assigned. Based on the assigned movement direction and their angular velocity, the movement pattern is generated. Their angular velocity is sampled uniformly between 8$^{\circ}$ and 15$^{\circ}$ per second. This angular velocity dictates how many 5$^{\circ}$ steps the talker traverses, resulting in discrete changes to the talker's spatial position (azimuth angle) over time.

This movement simulation was achieved by segmenting the 2.4-second clean speech utterance into short, time-varying blocks. For each block, a specific BRIR was selected from our pre-generated set, corresponding to the talker's angular position at that precise time within the trajectory. The selection process ensures that the sound for each segment is convolved with the BRIR associated with the talker's discrete spatial location during that segment. To minimize audible artifacts at the transitions between these segments, a short cross-fading window (5 ms) is applied between consecutive segments. This method generates spatially time-varying binaural signals with dynamic interaural cues, including Interaural Level Differences (ILDs) and Interaural Time Differences (ITDs), reflecting the speaker's changing location over time.

A total of 56,000 trajectory pairs were generated, split into 40,000 pairs for training, 10,000 for validation, and 6,000 for testing. The full code for trajectory simulation is available  \href{https://github.com/sayo20/Speech-Enhancement-for-Moving-Speakers-in-Classroom-Environment}{here}.

\subsubsection{Synthesis of the classroom speech data}
\label{datasets}
We used two distinct datasets for this study: (1) a classroom speech dataset with overlapping child–child and child–adult talker pairs, and (2) an adult-only dataset used to establish a baseline for generalization performance. In both datasets, spatialized two-talker mixtures were created by convolving clean speech utterances with BRIRs sampled along predefined movement trajectories (see Section~\ref{trajectories}). All waveforms were resampled to 16 kHz, and each utterance was 2.4 seconds in duration, matching the simulated trajectory length. See summary in Table \ref{tab:dataset_summary} and details below.

For the classroom dataset, we used the MyST corpus \cite{ward2019my} to source child utterances from speakers aged 8–11. We followed the official train/dev/test splits to ensure speaker disjointness across subsets. Low-quality segments labeled as \texttt{NO\_SIGNAL}, \texttt{SILENCE}, \texttt{DISCARD}, \texttt{SIDE\_SPEECH}, \texttt{ECHO}, or \texttt{BREATH} were removed. Adult speech used in mixed-age scenes was drawn from the LibriSpeech corpus \cite{panayotov2015librispeech}. We simulated two overlapping speech types: (1) two moving children, and (2) a moving child and a moving adult pair, both with and without background child babble. A diffuse babble noise field was generated using the CSLU corpus \cite{cole1998cslu}, with overlapping utterances (70\% overlap), spatialized using BRIRs from 3–8 random locations following \cite{francl2022deep}. 

The adult-only dataset was constructed solely from LibriSpeech utterances, following the mixture and movement framework described above. Two moving adult speakers were selected per mixture, and an additional adult babble stream was optionally added using utterances from the WSJ corpus\cite{paul92_icslp}. The adult babble diffuse noise field followed the same overlap and spatialization method described above for generating the children's babble noise field.

Across both datasets, we generated 40,000 training mixtures, 10,000 for validation, and 6,000 for testing. SNRs for the speech mixtures were sampled between 0 and 5 dB, while speech-to-babble SNRs ranged from –2.5 to 15 dB. 


\begin{table}[t]
\centering
\renewcommand{\arraystretch}{1.3}
\setlength{\tabcolsep}{4pt}

\begin{tabular}{
  p{1.9cm} 
  p{1.7cm} 
  p{2.3cm} 
  p{1.5cm} 
}
\hline
\textbf{Dataset} & \textbf{\thead[l]{Speaker \\ Types}} & \textbf{\thead[l]{Speech \\ Corpora}} & \textbf{\thead[l]{Babble \\ Corpora}} \\
\hline
Classroom
& \makecell[l]{Child–Child,\\ Child–Adult}
& \makecell[l]{MyST,\\ LibriSpeech }
& \makecell[l]{CSLU } \\
\hline
Adult-Only
& Adult–Adult
& LibriSpeech
& WSJ \\
\hline
\end{tabular}
\caption{Summary of the datasets created in this study. Both datasets contain spatialized two-talker mixtures with and without background babble noise.}
\label{tab:dataset_summary}
\end{table}

\begin{figure}[t]
    \centering
    \includegraphics[width=0.95\linewidth]{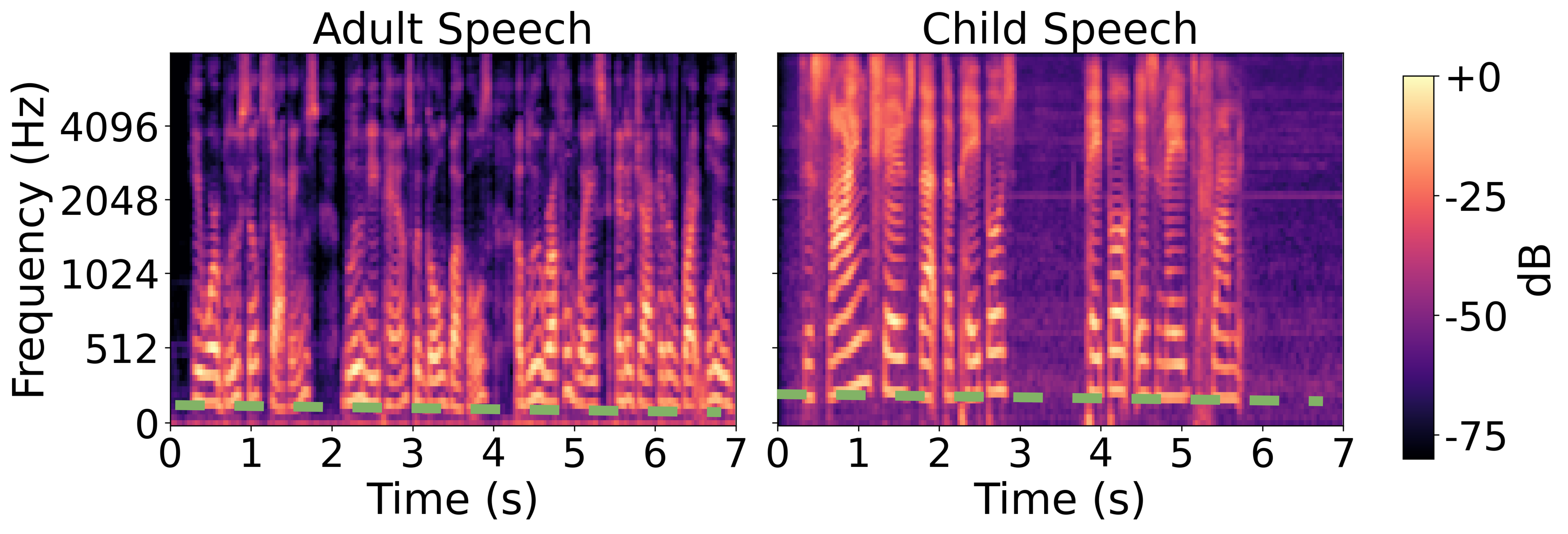}
    \caption{Spectrogram of utterances from an adult speaker (left) and a child speaker (right). The child’s speech shows higher fundamental frequency (F0) and greater pitch variation (visible in the more widely spaced and fluctuating harmonic structure). In contrast, the adult’s speech has a lower F0 and denser harmonic spacing, with energy more evenly distributed across a wider frequency range. The green dashed lines indicate estimated F0.}
    \label{fig:adult_child_spectrogram}
\end{figure}

To highlight acoustic differences between adult and child speech, Figure~\ref{fig:adult_child_spectrogram} presents annotated sample spectrograms of example utterances. As shown, the child's speech displays a higher fundamental frequency (F0), attributed to shorter vocal tracts and smaller vocal folds \cite{lee1999acoustics, vorperian2007vowel}. The child signal also exhibits narrower spectral bands and more pronounced pitch modulation, with harmonics less stable over time \cite{assmann2000time}. In contrast, the adult speech demonstrates broader spectral energy distributions and more stable harmonic structure  \cite{lee1999acoustics,assmann2000time}.
These differences in vocal tract characteristics, F0, and spectral shape are known to affect both perceptual intelligibility and the behavior of enhancement models. Furthermore, prior work has shown that children’s voices tend to be more spectrally similar to one another than adult voices \cite{lee1999acoustics,schuster2014speaker,gerosa2006analyzing}. This spectral overlap reduces the distinctiveness between speakers in a mixture, making source separation more challenging. In contrast, adult voices exhibit more inter-speaker spectral variability, providing models with stronger cues to differentiate speakers. This spectral similarity among children may help explain why models trained on adult speech could fail to generalize to child-child mixtures, where fewer acoustic cues are available for speaker discrimination.

\subsection{Binaural Speech Separation Model}
\label{model}

\begin{figure*}[ht]
    \centering
    \includegraphics[width=\textwidth]{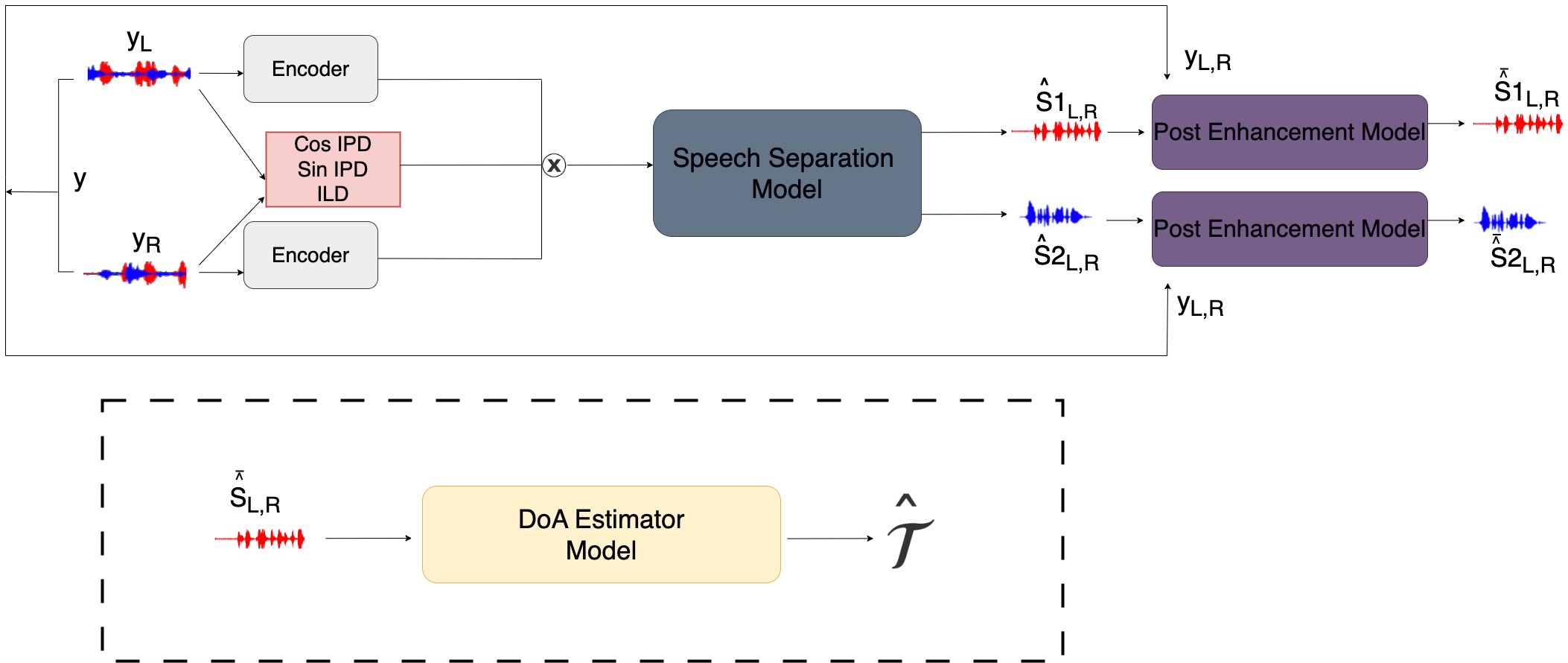} 
    \caption{Overall architecture of the proposed binaural speech separation and enhancement system \cite{han2021binaural}. The model processes left and right ear mixed signals ($y_L, y_R$) to output separated ($\hat{S}1_{L,R}$, $\hat{S}2_{L,R}$) and subsequently enhanced speech signals ($\overline{\hat{S}}1_{L,R}$, $\overline{\hat{S}}2_{L,R}$) for two sources. To evaluate whether spatial cues are preserved after enhancement, each enhanced binaural signal is further passed through a dedicated DoA estimation module, which predicts the estimated speaker trajectories ($\hat{\mathcal{T}}$). The DoA estimator is used exclusively for evaluation and is trained independently, ensuring it does not influence the separation or enhancement stages.}
    \label{fig:model_architecture}
\end{figure*}

We adopt the MIMO-TasNet-based architecture, as described in \cite{han2021binaural}, designed for real-time binaural speech separation. The model comprises two main components: a separation module and an enhancement module (see Figure~\ref{fig:model_architecture}). During evaluation, a Direction-of-Arrival (DoA) estimation module, which shares architectural similarities, operates on the enhanced outputs to classify the talker's trajectory. The entire model has 1.9 million parameters.

The separation module processes the binaural mixture, leveraging both spectral and spatial cues, including IPD and ILD, which are crucial for source localization. These cues, combined with learned representations from an encoder, are passed through a temporal convolutional network (TCN) to estimate time-domain source masks. The output consists of initial binaural waveform estimates corresponding to the two speakers in the mixture.

These initial estimates are then passed through the enhancement module, which shares an identical TCN-based architecture with the separation module but operates with independent parameters. This module takes as input both the noisy binaural speech mixture ($\mathbf{y}$) and the initial estimated speech signals ($\hat{\mathbf{s}}$). Its purpose is to refine and produce a cleaner version of the separated speech signals.

During training, we optimized the SNR loss between the estimated and reference signals, encouraging accurate waveform reconstruction while preserving spatial cues. The SNR is defined as:

\begin{equation} \text{SNR}(\mathbf{s}, \hat{\mathbf{s}}) = 10 \log_{10} \frac{||\mathbf{s}||^2}{||\mathbf{s} - \hat{\mathbf{s}}||^2} \end{equation}
where (s) and ($\hat{s}$) denote the reference and estimated speech signals, respectively.

To correctly match each separated speaker output to its corresponding reference signal, we applied permutation-invariant training (PIT) \cite{yu2017permutation}. Specifically, we minimized the loss over all possible speaker permutations:

\begin{equation}
\mathcal{L} = \min_{\pi \in P} \sum_{c=1}^{C} \left[ \text{SNR}(\mathbf{s}_{\pi(c)}^L, \hat{\mathbf{s}}c^L) + \text{SNR}(\mathbf{s}_{\pi(c)}^R, \hat{\mathbf{s}}_c^R) \right]
\end{equation}

where (P) is the set of all (C!) permutations, and C is the number of speakers. The same permutation ($\pi$) is applied to both left and right channels to maintain consistent speaker ordering across ears.

To evaluate whether the enhanced outputs successfully retained spatial information about the moving speakers, a DoA estimation module was employed. This DoA model was trained independently, only after the separation and enhancement modules had been completed, ensuring its sole purpose was to assess the preservation of spatial cues. Its architecture adopts the same TCN-based structure as the enhancement module but is trained as a classifier over discretized azimuth angles, following the approach described in \cite{han2021binaural} (see Fig.~\ref{fig:model_architecture}). The stereo signals from the enhancement output, $\hat{S}_{1,L,R}$ and $\hat{S}_{2,L,R}$, are encoded, passed through temporal convolutional layers, and mapped to per-frame DoA class probabilities. Let $\hat{\Theta}(t) \in \{ \theta_1, \ldots, \theta_K \}$ denote the predicted DoA class at time $t$. Each frame is labeled with a single ground-truth azimuth, and training is performed with a cross-entropy loss. During inference, these predictions are aggregated in fixed-length chunks, producing an estimated trajectory $\hat{\mathcal{T}} = \{\hat{\theta}_1, \ldots, \hat{\theta}_T\}$. By training the DoA estimator separately in this way, we ensure that its performance directly reflects how well the enhancement module preserved directional cues, rather than any confounding effects from joint optimization with the separation stage.

The model was implemented in the PyTorch framework \cite{paszke2019pytorch} and trained for 100 epochs using the Adam optimizer with an initial learning rate of 1×$10^{-3}$. An exponential learning rate scheduler (ExponentialLR) was applied with a decay factor of 0.98 every two epochs during the main training phase. During subsequent finetuning, the decay was applied every 5 epochs.

\subsection{Computational Experiments}
\label{sec:experiments}

All models in this study share the same MIMO-TasNet architecture described in Section~\ref{model}, but differ in the training data composition and noise conditions. Importantly, all experiments were rendered from the child-listener perspective using CHASAR BRIRs, ensuring that the simulated mixtures reflect the signals received by a child with bilateral hearing aids. To avoid ambiguity, models are named according to (i) the type of speech data used for training (adult vs. classroom) and (ii) whether babble noise was present during training (clean vs. babble).

\textit{Adult-Clean} and \textit{Adult-Babble} were trained exclusively on the adult dataset (adult–adult overlapping talkers), without and with background babble noise, respectively. These conditions serve as baselines and mirror common practice in prior work \cite{han2021binaural,gu2019neural, chen2018multi, wang2018combining, gu2020enhancing}, where models are trained primarily on adult speech. Evaluating these models on the classroom dataset allows us to assess their ability to generalize to child-dominant acoustic scenes.

\textit{Classroom-Clean} and \textit{Classroom-Babble} were trained entirely on the classroom dataset described in Section~\ref{datasets}, including child–child and child–adult overlapping talkers, again without and with background babble noise, respectively. These models reflect a child-aware training regime that explicitly targets the acoustic variability present in children’s speech and naturalistic classroom environments.

\textit{Finetune-Clean} and \textit{Finetune-Babble} were initialized from the corresponding Adult models and then finetuned using only 50\% of the classroom training data. These configurations test whether prior knowledge from adult speech separation can accelerate learning or improve generalization to children’s speech with reduced data requirements.

All models were evaluated on the classroom test set, referred to as \textit{Class-Clean } (With no babble) and \textit{Class-Babble} (With a babble background), allowing us to compare their performance in naturalistic school-like environments. We report the SNR and the SNR improvement (SNRi) to quantify the similarity between the separated output and the reference clean speech. We also report the estimated DoA error which measures the mean absolute angular deviation between the model's estimated talker position and the true position. Lower DoA is indicative of a more accurate spatial preservation. We further examine how training with or without babble noise influences robustness under noisy evaluation conditions.

\section{Results}

The results are organized to address three main questions. 
First, we ask whether models trained exclusively on adult talkers can generalize to child-listener classroom conditions. 
Within this analysis, we consider two dimensions of generalization: (i) noise robustness, comparing performance on clean versus babble mixtures, 
and (ii) voice-type effects, comparing adult–child versus child–child overlaps. 
Second, we evaluate whether training directly on classroom-specific data or fine-tuning an adult-trained model improves separation and localization performance in the classroom environment. 
Finally, we test the robustness of all models to additional factors, including the presence of babble during training and generalization to unseen source–listener distances.

\subsection{MIMO-Model trained on Adult-only Data}

A common issue in speech enhancement is the observable drop in performance when algorithms trained on adult voices are applied to children's speech. Existing studies typically use a single microphone. In the current study, we utilize recordings from two microphones placed on a bilateral hearing aid in order to leverage the inherent spatial cues that exist in the audio scenes, as well as introducing precomputed cues computed from both microphones. Our first objective is to determine if additional spatial and bilateral cues mitigate the performance drop typically observed when using adult-trained models on children's speech.

\subsubsection{Noise generalization}

Table~\ref{tab:adult_generalization} summarizes how the model trained on adult-only data with and without babble performs on two different test sets. 

When the model trained on the Adult-Clean data, is evaluated on the adult-clean test data, we observe an SNRi of 11.37 dB. When the same model is evaluated on the class-clean test data, we observe a SNRi of 10.81 dB. While this reduction is small, it was statistically significant (Mann–Whitney $U = 7.80 \times 10^{7}$, $p < 10^{-29}$, $r = -0.084$). Furthermore, we see that the Adult-Clean model was better able to track the movement in the adult-clean test data (DoA error= 3.82°) in comparison to the movements in the class-clean test data (DoA error = 4.90°; Mann–Whitney $U = 6.3 \times 10^{7}$, $p < 10^{-61}$, $r = 0.123$).

When the model trained on the Adult-Babble data, is evaluated on the adult-babble test data, we observe an SNRi of 9.92 dB. When the same model is evaluated on the class-babble test data, we observe a SNRi of 8.50 dB. The difference observed was also statically significant (Mann–Whitney $U = 8.91 \times 10^{7}$, $p < 10^{-221}$, $r = -0.237$). Although overall speech separation performance (SNRi) dropped significantly, indicating a clear domain shift challenge, the DoA error demonstrated a conflicting trend. The model showed a statistically significant, albeit small, improvement in tracking movements in the Class-Babble dataset ($10.81°$ vs. $11.37°$; Mann–Whitney $U = 8.13 \times 10^{7}$, $p < 10^{-68}$, $r = -0.13$). This result suggests that while the spectral characteristics of the child babble interfere with mask estimation (reducing SNRi), the spatial cues required for DoA estimation remain robust, or are even slightly clearer, in the simulated classroom environment.

Taken together, we observed that introducing background babble leads to a higher mismatch, likely because the properties of babble noise differ between adult and children's voices.

\subsubsection{Voice-type generalization}

As described in section \ref{datasets}, our Classroom dataset is composed of adult-child talker pairs and child-child talker pairs. We evaluate how the Adult-Clean and Adult-Babble models perform on these two different speaker configurations (see Table \ref{tab:Adult_model_performance}.

We observed a performance drop across both adult-child and child-child pairings in both clean and babble conditions. For the Adult-Clean model, SNRi decreased from 11.38 dB in adult–child mixtures to 10.25 dB in child–child mixtures (Mann–Whitney $U = 2.03 \times 10^{7}$, $p < 10^{-32}$, $r = -0.125$). Similarly, the model had a better DoA error  of 4.56° in adult–child mixtures than 5.24° in child–child mixtures (Mann–Whitney $U = 1.63 \times 10^{7}$, $p < 10^{-19}$, $r = 0.092$). This reduction, even in the absence of background noise, shows that the spectral mismatch between children's voices still makes separation more difficult for adult-trained models.

The same pattern is observed under babble noise. For the Adult-Babble model, SNRi dropped from 8.92 dB (adult–child) to 8.08 dB (child–child; Mann–Whitney $U = 2.02 \times 10^{7}$, $p < 10^{-29}$, $r = -0.120$). However, with the introduction of children babble, the model performed similarly in the localization of the adult–child pairs (DoA error = $10.59°$) and the child–child pairs (DoA error = $11.03°$), as the difference was not statistically significant (Mann–Whitney $U = 1.83 \times 10^{7}$, $p = 0.06$, $r = -0.20$ ).

Taken together we observe that when both talkers are children, the combined effects of spectral similarity and environmental noise impose a significant challenge on adult-trained models. Nevertheless, we see that the model still maintains its ability to separate and track the child talkers when it can leverage the binaural audio and spatial cues.


\begin{table}[h]
\renewcommand{\arraystretch}{1.2}
\setlength{\tabcolsep}{6pt}
\centering

\begin{tabular}{|l|c|c|}
\hline
\multirow{2}{*}{\textbf{Test Data}} & \multicolumn{2}{c|}{\textbf{Adult-Clean Model}} \\
\cline{2-3}
 & \textbf{SNRi (dB) $\uparrow$} & \textbf{DoA ($^\circ$) $\downarrow$} \\
\hline
Adult-Clean & 11.37 & 3.82 \\
Class-Clean & 10.81 & 4.90 \\
\hline
\end{tabular}

\medskip

\begin{tabular}{|l|c|c|}
\hline
\multirow{2}{*}{\textbf{Test Data}} & \multicolumn{2}{c|}{\textbf{Adult-Babble Model}} \\
\cline{2-3}
 & \textbf{SNRi (dB) $\uparrow$} & \textbf{DoA ($^\circ$) $\downarrow$}\\
\hline
Adult-Babble & 9.92 & 11.37 \\
Class-Babble & 8.50 & 10.81 \\
\hline
\end{tabular}

\caption{Performance of adult-trained models on in-domain (Adult test sets) and out-of-domain (Classroom test sets). 
The Adult-Clean model was trained without babble noise and tested on Adult-Clean and Classroom-Clean sets, while the Adult-Babble model was trained with babble noise and tested on Adult-Babble and Classroom-Babble sets.}
\label{tab:adult_generalization}
\end{table}


\begin{table}[h]
\renewcommand{\arraystretch}{1.2}
\setlength{\tabcolsep}{6pt}
\centering
\begin{tabular}{|l|c|c|}
\hline
\multirow{2}{*}{\textbf{Overlap Type}} & 
\multicolumn{2}{c|}{\textbf{Adult-Clean Model}} \\
\cline{2-3}
 & \textbf{SNRi (dB) $\uparrow$} & \textbf{DoA (°) $\downarrow$} \\
\hline
Adult–Child & 11.38 & 4.56 \\
Child–Child & 10.25 & 5.24 \\
\hline
\end{tabular}

\vspace{0.7em}

\begin{tabular}{|l|c|c|}
\hline
\multirow{2}{*}{\textbf{Overlap Type}} & 
\multicolumn{2}{c|}{\textbf{Adult-Babble Model}} \\
\cline{2-3}
 & \textbf{SNRi (dB) $\uparrow$} & \textbf{DoA (°) $\downarrow$} \\
\hline
Adult–Child & 8.92 & 10.59 \\
Child–Child & 8.08 & 11.03 \\
\hline
\end{tabular}
\caption{Performance of Adult-Clean and Adult-Babble models on classroom mixtures, partitioned by overlap type (adult–child vs.\ child–child). 
The Adult-Clean model is evaluated on the Classroom-Clean test set, while the Adult-Babble model is evaluated on the Classroom-Babble test set.}
\label{tab:Adult_model_performance}
\end{table}

\subsection{MIMO-Model trained on Adult Data vs MIMO-Model trained on Classroom Data}


In this section we test whether direct exposure to domain-specific data improves performance and robustness in the classroom, especially with child-child talkers. To investigate this, we evaluated the effect of training the model directly on the Classroom dataset.

\subsubsection{Performance in Clean Classroom Conditions}

We first compared the Adult-Clean and Classroom-Clean models on the Class-Clean test data (Table~\ref{tab:clean_babble_split}, Fig.~\ref{fig:mimo_interactions}A). 
The two models achieved nearly identical overall performance (SNRi = 10.81 vs.\ 10.80 dB, DoA = 4.90° vs.\  4.68°). This suggests that in the absence of babble, the spatial cues are sufficient to allow adult-trained models to generalize remarkably well to classroom-related speech mixtures.

However, when we zoom into the different talker pairs in the Class-clean data, a consistent gap remained between the model performance on adult–child and child–child talkers. The Adult-clean model had a SNRi of 11.36 dB and DoA error of 4.56° on the adult-child pair and 10.25 dB and DoA error of 5.24°  on the child-child pair (drop in SNRi =  1.11 dB , increase in doa = 0.68°; Mann–Whitney $U=2.03 \times 10^{7}$, $p<10^{-32}$, $r=-0.125$, FDR-corrected). While the Classroom-clean Model had a SNRi of 11.25 dB and DoA error of 4.50° on the adult-child pair and 10.37 dB and doa of 4.85° on the child-child pair (drop in SNRi = 0.88 dB, increase in DoA error = 0.68°; Mann–Whitney $U=1.96 \times 10^{7}$, $p<10^{-16}$, $r=-0.089$, FDR-corrected).

Taken together, we observed that while domain-specific training provides a slight improvement for the most challenging case (child–child mixtures), it does not offer a substantial advantage over the Adult-clean model in clean classroom conditions.



\subsubsection{Performance in Noisy Classroom Conditions}

\begin{figure}[h]
    \centering
    \includegraphics[width=\linewidth]{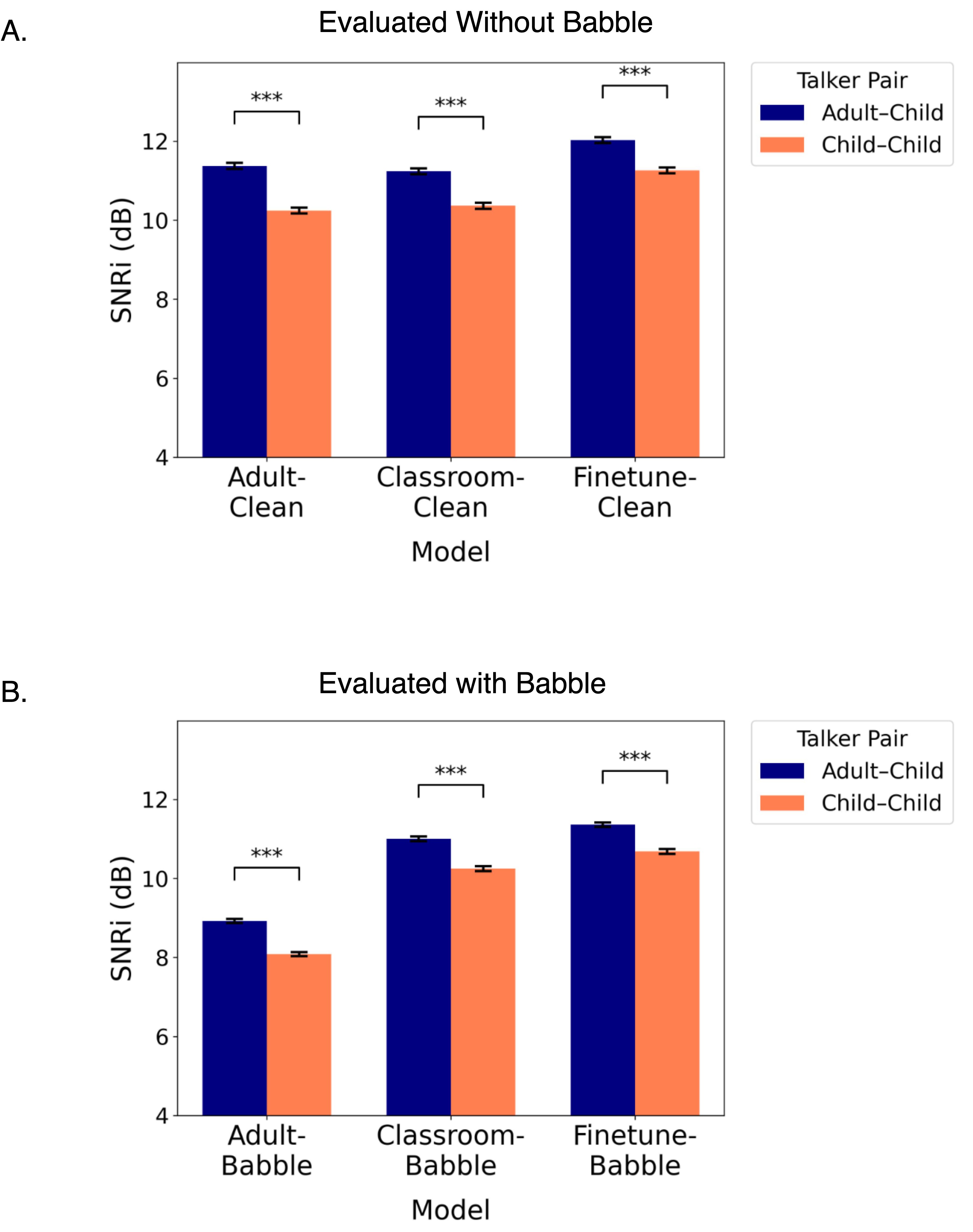} 
    \caption{Speech separation performance of MIMO models trained on different datasets and evaluated on classroom conditions under varying noise conditions. Bar plots show the mean signal-to-noise ratio improvement (SNRi) for the MIMO model trained on the different data splits, Adult, Class, and Finetuned, evaluated on the Class dataset involving either Adult–Child (blue) or Child–Child (orange) talker pairs.
    (A) Performance of models trained and evaluated in babble-free reverberant classroom conditions.
    (B) Performance of models trained and evaluated with background babble in reverberant classroom conditions. Error bars indicate the standard error of the mean (SEM). Asterisks (***) denote statistically significant differences across conditions within each model (p  $<$ 0.001, Mann-Whitney U test).}
    \label{fig:mimo_interactions}
\end{figure}

In babble conditions, domain-specific training conferred a clearer advantage (Table~\ref{tab:clean_babble_split}, Fig.~\ref{fig:mimo_interactions}B). 

When evaluated on the Class-Babble test data, the Classroom-Babble model achieved an SNRi of 10.62 dB and a DoA of 6.78°, which is a substantial improvement over the Adult-Babble model's performance (SNRi: 8.50 dB, DoA error: 10.81°; Mann–Whitney $U=9.59 \times 10^{7}$, $p<10^{-60}$, $r=-0.28$, FDR-corrected). This demonstrates that direct exposure to the specific acoustic properties of babble noise and children's voices is critical for robust performance in noisy environments.

The Classroom-Babble model also perform better on the child-child talker pairs, with an SNRi of 10.25 dB, and DoA error of 6.94°. While the Adult-Babble model achieved an SNRi of 8.08 dB, and DoA error of 10.78° on the child-child talker pairs. Furthermore, we observed a smaller performance drop between the talker pairs for the Classroom-Babble model. The Adult-Babble model showed a 0.84 dB drop in SNRi between adult-child and child-child talkers (Mann–Whitney $U=1.84 \times 10^{7}$, $p<10^{-30}$, $r=-0.12$, FDR-corrected), while the drop in the Classroom-Babble model was 0.76 dB (Mann–Whitney $U=1.96 \times 10^{7}$, $p<10^{-17}$, $r=-0.01$, FDR-corrected) (see Fig.~\ref{fig:mimo_interactions}B).

Taken together, these results show that while using binaural audio and providing spatial cues can alleviate a significant portion of the performance loss seen with adult-trained models, they still suffer in noisy environment. Direct exposure to the target domain through classroom-specific training provides robustness in noisy conditions.

\begin{table*}[h]
\renewcommand{\arraystretch}{1.2}
\setlength{\tabcolsep}{6pt}
\centering
\subfloat[\textbf{Models trained without babble noise}]{
\begin{tabular}{|l|cc|cc|cc|}
\hline
\multirow{2}{*}{\textbf{Test Data}} &
\multicolumn{2}{c|}{\textbf{Adult-Clean}} &
\multicolumn{2}{c|}{\textbf{Classroom-Clean}} &
\multicolumn{2}{c|}{\textbf{Finetune-Clean}} \\
\cline{2-7}
 & \textbf{SNRi (dB) $\uparrow$} & \textbf{DoA (°) $\downarrow$} 
 & \textbf{SNRi (dB) $\uparrow$} & \textbf{DoA (°) $\downarrow$} 
 & \textbf{SNRi (dB) $\uparrow$} & \textbf{DoA (°) $\downarrow$} \\
\hline
Class-Clean & 10.81 & 4.90 & 10.80 & 4.68 & 11.65 & 4.11 \\
Class-Babble & 7.77 & 13.85 & 8.01 & 13.05 & 8.00 & 12.58 \\
\hline
\end{tabular}
}

\vspace{1em}

\subfloat[\textbf{Models trained with babble noise}]{
\begin{tabular}{|l|cc|cc|cc|}
\hline
\multirow{2}{*}{\textbf{Test Data}} &
\multicolumn{2}{c|}{\textbf{Adult-Babble}} &
\multicolumn{2}{c|}{\textbf{Classroom-Babble}} &
\multicolumn{2}{c|}{\textbf{Finetuned-Babble}} \\
\cline{2-7}
& \textbf{SNRi (dB) $\uparrow$} & \textbf{DoA (°) $\downarrow$} 
 & \textbf{SNRi (dB) $\uparrow$} & \textbf{DoA (°) $\downarrow$} 
 & \textbf{SNRi (dB) $\uparrow$} & \textbf{DoA (°) $\downarrow$}  \\
\hline
Class-Clean & 9.76 & 6.36 & 10.48 & 5.74 & 11.05 & 4.99 \\
Class-Babble & 8.50 & 10.81 & 10.62 & 6.78 & 11.02 & 6.55 \\
\hline
\end{tabular}
}
\caption{Performance of models trained without (a) and with (b) babble noise, evaluated on classroom mixtures with and without babble. All models were trained and evaluated from the child-listener perspective.}
\label{tab:clean_babble_split}
\end{table*}

\subsection{MIMO-Model trained on Adult Data Finetuned on Classroom Dataset }

While full-domain training on classroom data is beneficial, particularly in noisy conditions, acquiring a large, high-quality children's speech corpus is a well-known challenge. This makes children's speech a low-resource domain. Given that the adult-trained model already shows strong generalization in clean conditions due to spatial cues, it suggests that the model's foundational knowledge of speech separation is transferable. Therefore, we investigate a more data-efficient approach: fine-tuning the robust adult-trained model with a limited amount of classroom data. This allows us to determine whether we can achieve performance comparable to a fully trained classroom model without the need for a large, specialized dataset.

\subsubsection{Comparison of Classroom-clean and Finetuned-clean Models}

The Finetuned-Clean model (see Table \ref{tab:clean_babble_split}) outperformed the Classroom-Clean model across both metrics, achieving an SNRi of 11.65 dB (vs. 10.80 dB) and a lower DoA error of 4.11° (vs. 4.68°), despite being trained on only half the amount of classroom scenes. This difference was statistically significant (Mann–Whitney $U=6.39 \times 10^{7}$, $p<10^{-50}$, $r=-0.11$, FDR-corrected).

Furthermore, fine-tuning showed a slightly smaller drop in performance between the adult-child and the child-child talker pairs. The Finetuned-Clean model exhibited an SNRi drop of 0.77 dB between the adult-child and child-child pairings (Mann–Whitney $U=1.9 \times 10^{6}$, $p<10^{-15}$, $r=-0.08$, FDR-corrected), compared to the 0.84 dB drop observed in the Classroom-Clean model (see Fig.\ref{fig:mimo_interactions}A). Fine-tuning not only improved overall performance but also enhanced the model's robustness to the most difficult speaker configurations.

These findings demonstrate that fine-tuning effectively transfers spectral priors learned from adult speech representations to the children's classroom domain. This strategy is particularly effective in clean conditions, as it leverages the model's existing knowledge of speech acoustics without the added noise complexity.

\subsubsection{Comparison of Classroom-babble and Finetuned-babble Models}

On the Class-Babble test data (see Table \ref{tab:clean_babble_split} b), the Finetuned-Babble model achieved an SNRi of 11.02 dB and a DoA error of 6.55°, which is a significant improvement over the Classroom-Babble model's 10.62 dB SNRi and 6.78° DoA error (Mann–Whitney $U=6.79 \times 10^{7}$, $p<10^{-14}$, $r=-0.06$, FDR-corrected). As with the Finetuned-Clean model, the Finetuned-Babble model was trained with only half of the Classroom scenes, as was used in the Classroom-Babble model.

The Finetuned-Babble model also slightly performed better at handling the child-child talker pairs (SNRi = 10.68 dB , DoA error = 6.55° ), outperforming the fully-trained Classroom-Babble model (SNRi = 10.25 dB , DoA error = 6.94°). The Finetuned-Babble model had a small  0.68 dB SNRi drop between the adult-child and child-child talker pair (Mann–Whitney $U=1.9 \times 10^{6}$, $p<10^{-14}$, $r=-0.08$, FDR-corrected), compared to the 0.76 dB drop observed in the Classroom-Babble model (see Fig.\ref{fig:mimo_interactions}B).

These results collectively demonstrate that fine-tuning an adult-trained model with a limited amount of domain-specific data is not just a viable alternative but could be a better strategy for speech separation in a noisy classroom. 

\subsection{Robustness to Unseen Training Conditions}
To assess the model's robustness and generalization capabilities, we evaluated its performance on acoustic conditions on which it was not explicitly trained. We investigate two key scenarios: (i) Noise generalization: We first examine how models trained with and without babble noise perform when exposed to the opposite condition. (ii) Distance generalization: We then take the worst and best performing model trained on the noisy condition (Adult-Babble, Finetune-Babble) and test its ability to generalize to unseen talker distances. 

\subsubsection{Noise generalization; Impact of Babble During Training}

\begin{figure}[h]
    \centering
    \includegraphics[width=\linewidth]{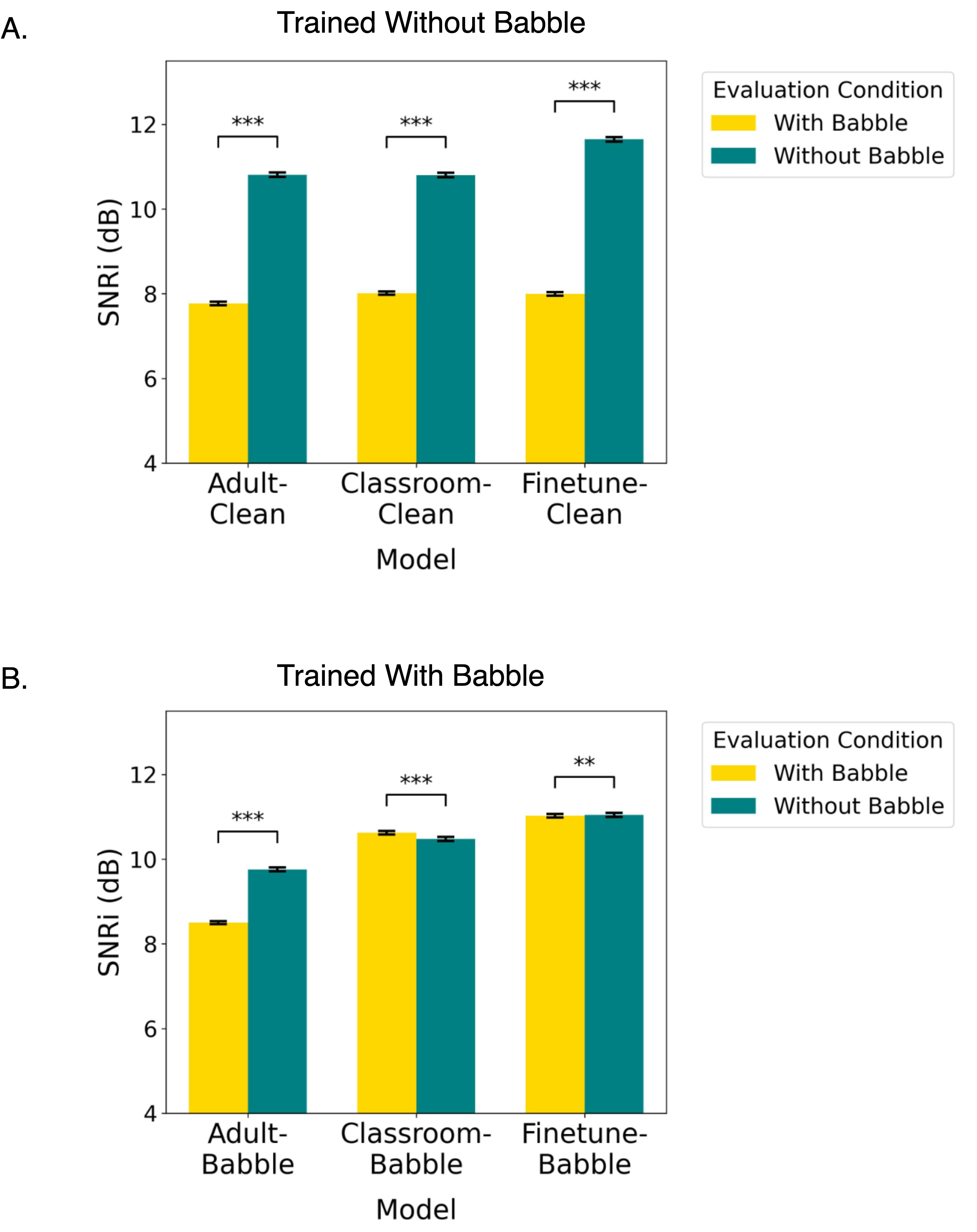} 
    \caption{SNR performance of three models under clean and noisy evaluation conditions. (A) Models trained without babble noise. (B) Models trained with babble noise. All models were evaluated on classroom data with and without background babble.
    Teal bars represent evaluations without babble; Yellow bars represent evaluations with babble. Error bars indicate the standard error of the mean (SEM). Asterisks (***) denote statistically significant differences (p $<$ 0.001, Mann-Whitney U test).}
    \label{fig:babble_effect}
\end{figure}

The results (See Table. \ref{tab:clean_babble_split} and Fig.~\ref{fig:babble_effect}) showed a clear asymmetry in model's robustness. Models trained without babble noise, Adult-Clean, Classroom-Clean, and Finetuned-Clean demonstrated significant performance degradation when tested on noisy mixtures (Table. \ref{tab:clean_babble_split} a). For example, Finetuned-Clean's SNRi dropped sharply from an SNRi of 11.65 dB and a DoA error of 4.11° on clean mixtures to  an SNRi of only 8.00 dB and a DoA error of 12.58° on noisy mixtures (see Fig.~\ref{fig:babble_effect} A; Mann–Whitney $U=4.3 \times 10^{7}$, $p<10^{-60}$, $r=-0.02$, FDR-corrected), highlighting the vulnerability of the clean-trained models to unseen noise, which is common in a classroom environment.

In contrast, models trained with babble noise showed better generalization. These models not only performed robustly in noisy conditions but also transferred well to clean environments. The Finetuned-Babble model demonstrated the strongest generalization. It maintained a high SNRi of 11.05 dB and a DoA error of 4.99° in clean conditions, and an SNRi of 11.02 dB and  DoA error of 6.55° in noisy conditions. The Classroom-Babble and Adult-Babble models also showed good generalization, with performance on clean and noisy data being closely matched (10.48 dB and 10.62 dB for Classroom-Babble; 9.76 dB and 8.50 dB for Adult-Babble). 

\subsubsection{Generalization to Unseen Talker Distances}
To assess the models' ability to track talkers in different spatial configurations, we evaluated the Adult-Babble and Finetuned-Babble models on the Class-Babble test set with talkers at increased distances of 1.5 m and 2.0 m, as they were only trained on 1.0 m configurations. This setup was made to simulate some variability in the speaker–listener distances.

\begin{figure}[h]
    \centering
    \includegraphics[width=\linewidth]{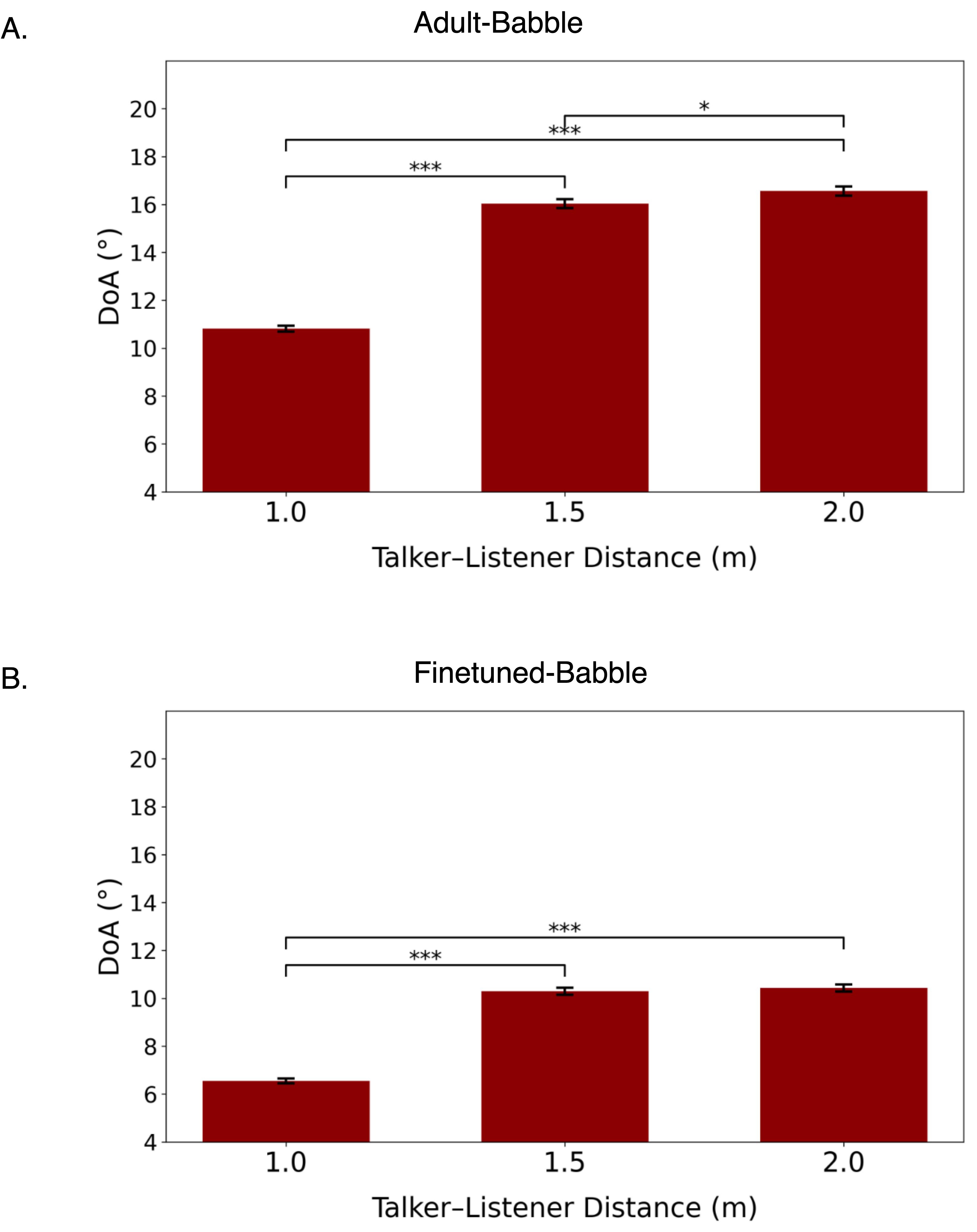} 
    \caption{Average direction-of-arrival error (DoA) achieved by the Adult-Babble (A) and Finetuned-Babble (B) when evaluated on the Class-Babble test set at three talker–listener distances: 1.0~m, 1.5~m, and 2.0~m. Error bars indicate the standard error of the mean (SEM). Asterisks denote statistically significant differences between distances (Mann-Whitney U, FDR-corrected; * $p < 0.05$, *** $p < 0.001$).}
    \label{fig:distance_effect}
\end{figure}

As shown in Fig.~\ref{fig:distance_effect}, both the Adult-Babble and Finetuned-Babble models showed a decline in performance as talker distance increased. However, the Finetuned-Babble model demonstrated more robustness to the unseen talker distances. The Finetuned-Babble model's DoA error increased by 3.74° between 1.0 m and 1.5 m, and by 3.89° between 1.0 m and 2.0 m. In contrast, the Adult-Babble model's error increased by 5.22° and 5.75° respectively.


\section{Discussion}

This study explored how DNN-based speech separation can be adapted to classroom environment and children's speech to provide effective speech enhancement for hearing-impaired children. Our approach leverages binaural audios, which are simulated yet naturalistic acoustic scenes with two moving talkers. Our findings demonstrate that while the model trained solely on binaural adult speech recordings can generalize to children's speech in quiet conditions, its performance significantly degrades in the presence of children's babble noise. Crucially, we show that training on the classroom dataset and, even more effectively, fine-tuning a pre-trained model are essential strategies for achieving robust performance in noisy classroom settings. The success of these adaptation methods has direct and important implications for developing adaptive hearing aids and assistive listening devices that can better serve children in dynamic educational settings.

Prior work has consistently shown that models trained on adult speech generalize poorly to children's speech, primarily due to key spectral differences, such as higher fundamental frequencies and less stable harmonics \cite{potamianos1997automatic, shivakumar2020transfer,attia2025cpt}. These studies, however, are almost exclusively based on monaural recordings, where separation relies heavily on spectral distinctiveness of the talkers, e.g, the disimilarity between adult and female talkers \cite{wang2020study, wang2018progressive}. We hypothesized that a binaural approach could mitigate this challenge. Based on our previous work \cite{olalere2025leveraging}, which showed that a binaural model can effectively leverage spatial cues (such as interaural time and level differences) to resolve spectrally similar adult talkers, we hypothesized that this spatial mechanism could be robust enough to overcome the spectral challenges inherent in child-talker separation. Our results confirm that, in clean conditions, the performance drop between the Adult-Only model (trained exclusively on adult pairs) and the Classroom model (trained on mixed adult-child and child-child pairs) is minimal. This finding is significant as it suggests that, in the absence of noise, the spatial cues alone are sufficient to reduce the spectral mismatch barrier observed in speech task using monaural recordings.

This spatial advantage leveraged by the Adult-Clean model, however, diminishes in the presence of babble noise, which is a main characteristic of classrooms. Our results suggest that the acoustic nature of the noise itself, rather than the voice mismatch, is the primary barrier. The Adult-Babble model, despite being trained on adult babble, still failed to generalize to the unique characteristics of children's babble. This aligns with prior work on children's listening challenges \cite{howard2010listening} and underscores the importance of a domain-specific noise component in building robust models. To address the resulting need for domain-specific noise training in this low-resource domain, we investigated fine-tuning on classroom data as an efficient strategy \cite{shivakumar2020transfer, karunanayake2019transfer}. The Finetuned-Babble model performed better than the Adult-Babble model on the Class data. This demonstrate that the spectral priors learned from adult speech and adult babble provide a strong foundation for rapid adaptation to children's speech in both clean and noisy environments, using only 50\% of the target data. This hybrid approach is valuable for low-resource domains and also offers practical benefits, such as reducing training time from 17 hours to just 8 hours in our case.
We therefore recommend prioritizing resource-efficient adaptation methods, such as fine-tuning with limited additional data, while in parallel expanding high-quality datasets of children’s speech to enable effective adaptation. While our fine-tuning strategy was simple, more sophisticated methods, such as discriminative layer-wise tuning or parameter-efficient transfer methods \cite{howard2018universal,hu2022lora}, could unlock even greater improvements.

Beyond speech separation, we show that our models are able to preserve the spatial location of the talkers in noise, as evidenced by the reduction of the DoA error. Accurate sound localization can help the listeners orient themselves and focus in complex, multi-talker scenes \cite{bronkhorst2000cocktail,litovsky2006bilateral}. Our results show that models trained on the classroom data, especially the finetuned model, achieved significantly lower DoA error, even with babble noise present. This demonstrates that these models are not just separating speech from noise, but are also learning to preserve the spatial information of the original talker. Furthermore, the performance of the finetuned model in this regard suggests that spatial cue preservation is a learned feature that can benefit from both general acoustic priors (from adult training) and targeted domain-specific examples (from classroom data). The capability of this model can be further developed for hearing assistive technologies, for example, to provide accurate, low-error direction estimates to steer neural beamformers on hearing aids. 

Finally, our findings on the model's robustness to unseen conditions and talker distances are particularly encouraging as a step towards real-world deployment. We see a consistent pattern of babble-trained models outperforming their clean-trained counterparts in both clean and noisy conditions. The Adult-clean, Classroom-clean , and Finetuned-clean's model's ability to track the talkers movement particularly degrades once babble noise is present in the acoustic scene.
Furthermore, the asymmetry in performance where models trained with noise generalize well to clean conditions but not vice-versa mirrors established principles in machine learning, suggesting that the exposure to babble noise acts as a form of data augmentation that forces the model to learn more generalized and robust features\cite{hershey2016deep,botinhao2016investigating}. Also, the ability of the Finetuned-Babble model to separate and track talkers at unseen distances (1.5 m and 2.0 m) is a good indicator of its generalization capabilities. While a performance decline is expected with increasing distance due to lower signal energy and a reduced direct-to-reverberant energy ratio \cite{kuttruff2016room}, the model still maintained functional separation and spatial tracking. This suggests that the model had internalized fundamental, scale-invariant spatial patterns that allowed it to generalize beyond its training configuration. Moreover, incorporating a range of talker distances during training could enhance the model’s ability to generalize even better to unseen distance configurations that could occur in classroom environments.


This study is not without limitations. Our simulations, while aiming for acoustic realism, are not based on real classroom recordings. We also focused on simplified two-speaker mixtures and fixed talker-to-listener distances during movement. Future work should involve developing datasets that account for more complex scenarios, including multiple simultaneous talkers and varying speaker distances. Additionally, while our standard fine-tuning approach proved effective, more advanced domain adaptation techniques can be explored for even better performance. Furthermore, while objective metrics like SNRi do not translate linearly to intelligibility gains observed in human listening tests, prior studies have shown that higher SNRi generally correspond to better speech understanding and reduced listening effort in hearing-impaired listeners \cite{thoidis2024using, pichora2016hearing, brons2014effects}. In this context, the consistent SNRi gains and spatial cue preservation observed across our fine-tuned models can be interpreted as a strong indication that similar perceptual benefits—such as improved speech clarity and ease of listening—would likely occur if the processed signals were delivered through a hearing device. As such, field validation and perceptual testing are essential next steps to confirm the model's robustness in real-world settings. Despite these limitations, our findings provide strong evidence that the use of binaural spatial cues enables speech separation models to better bridge the gap when adult-trained systems are applied in classroom environments. At the same time, the results highlight the importance of domain-specific classroom data to capture the unique noise characteristics of these settings.

\section{Conclusions}
\label{conclusion}

This study demonstrates that binaural spatial cues can substantially reduce the performance gap when adult-trained speech separation models are applied to children’s speech in classroom environments. While adult-trained models generalize reasonably well in clean conditions, their performance degrades sharply in the presence of realistic children’s babble noise. We showed that domain-specific classroom training improves noise robustness and that finetuning adult-trained models with limited classroom data is even more effective, offering both higher performance and reduced training costs. Crucially, finetuned models also preserved spatial localization cues, enabling accurate talker tracking under challenging, noisy conditions.

\section{Acknowledgements}
\noindent This work is part of the INTENSE consortium, which has received funding from the NWO Cross-over Grant No. 17619. We also acknowledge the use of AI-assisted tools such as Grammarly and Google Gemini for grammar checking and enhancing the clarity of the manuscript. We take full responsibility for the work.

\bibliographystyle{ieeetr}
\bibliography{Main}

\end{document}